# Enhanced LIBS Emission Using Laser Beam Splitting: Interacting Multi-Plume Plasma Dynamics

Girum Abebe Beyene*, Inam Mirza, Jijil JJ Nivas, Oleksandr Gatsa, Nadezhda M. Bulgakova*, Alexander V. Bulgakov

FZU - Institute of Physics of the Czech Academy of Sciences, Na Slovance 1999/2, 182 00, Prague 8, Czech Republic

*Authors to whom correspondence should be addressed.
E-mail: girum.beyene@fzu.cz and nadezhda.bulgakova@fzu.cz

**Abstract**. The optical emission in laser-induced breakdown spectroscopy (LIBS) is governed by the spatial intensity distribution of the incident laser beam, which influences plasma formation and evolution. Beam shaping therefore offers a route to control plasma dynamics and emission yield; however, its effects in LIBS remain insufficiently quantified, particularly under conditions relevant to compact instrumentation. In this work, a diffractive optical element (DOE) is used to transform a Gaussian beam into a 2×2 array, producing simultaneously expanding, co-propagating ablation plumes that interact during expansion. Plasma evolution from Cu and Si targets is investigated in vacuum using a Nd:YAG laser (1064 nm, 5 ns, 10 J $cm^{-2}$), combining time-resolved imaging with optical emission spectroscopy. The multi-spot configuration results in enhanced emission intensity compared to single-spot irradiation, with increases of ~9 for Si and ~3 for Cu. The observed enhancement is attributed to plume–plume interaction effects that modify plasma density and emission characteristics. These results demonstrate that DOE-based beam shaping provides an effective and technically simple approach to increasing the LIBS signal without additional system complexity.

## 1. Introduction

Laser-induced breakdown spectroscopy (LIBS) is a versatile analytical technique for rapid, remote, and *in situ* elemental analysis, with applications across a wide range of scientific and technological fields [1–5]. It offers significant advantages over conventional techniques in harsh or inaccessible environments, including space exploration [6–10], deep-sea studies [11], monitoring plasma-wall interaction in nuclear fusion reactors [12], geological and environmental analysis [13–16], and industrial quality control [17,18]. Recent developments in data processing, particularly the application of machine learning [15,16], advances in miniaturized instrumentation [19], and hybrid approaches combining LIBS with Raman spectroscopy [16], laser-induced fluorescence (LIF) [20], or mass spectrometry (MS) [21], further highlight the growing capabilities of LIBS. Despite these advances, widespread industrial adoption of LIBS remains limited by insufficient repeatability, low sensitivity, and reduced analytical accuracy. Among the key factors governing LIBS performance is the signal-to-noise ratio, the improvement of which is essential for enhancing detection sensitivity.

A variety of approaches have been proposed to increase LIBS signal intensity. These methods can be broadly classified into into three major categories [22]: multi-pulse excitation schemes (e.g., double- or triple-pulse LIBS) [23–25]; sample modification techniques such as nanoparticle assisted LIBS [26,27]; and the use of auxiliary fields or confinement methods, including electric [28], magnetic [29], and microwave [30] fields, spark assistance [31], and spatial confinement [32]. While these approaches can significantly enhance emission signals under controlled conditions, they often require additional hardware, increase system complexity, and reduce robustness. Such limitations pose challenges for practical implementation, particularly in industrial environments where compactness, stability, and ease of operation are critical [33]. For example, double-pulse LIBS requires precise synchronization and optimization of parameters such as inter-pulse delay, relative pulse energies, and temporal ordering [23,24]. Achieving optimal performance is non-trivial due to the strong dependence of transient plasma properties, including electron temperature and density, which are inherently dynamic and spatially non-uniform.

An alternative approach to improving LIBS performance is spatial beam shaping, in which non-Gaussian beam profiles – such as Bessel beams, vortex beams carrying orbital angular momentum, and flat-top beams – are employed to control energy deposition pathways and subsequent plasma evolution [33–35]. Diffractive optical elements (DOEs) have recently been used to control laser–matter interaction in applications such as periodic surface structuring [36] and pulsed laser ablation in liquid [37,38]. In this work, we extend this concept to nanosecond laser excitation for LIBS by employing a DOE to reshape the spatial intensity distribution of the incident beam. Specifically, a Gaussian beam is split into a 2×2 array, generating four ablation plumes that expand simultaneously and interact during their evolution. To the best our knowledge, the use of DOE-based multi-spot irradiation to produce interacting co-propagating plumes for LIBS signal enhancement has not been previously reported.

Modern optical fabrication techniques enable the production of high-efficiency DOEs capable of generating one- or two-dimensional beam arrays with well-defined intensity distributions [39]. These elements offer several practical advantages, including compact form factor, high diffraction efficiency, robustness under high-power operation, and tolerance to minor misalignment. In the context of nanosecond LIBS, beam splitting may also mitigate plasma shielding effects commonly observed with single Gaussian beams, where increased plasma absorption limits energy coupling to the target at higher fluences. However, the use of DOEs is not without limitations. In particular, the fixed angular separation between diffracted beams constrains the achievable spacing between the irradiation spots on the target.

In this work, we investigate the use of DOE-based beam splitting as a simple and effective approach for enhancing LIBS emission. Plasma formation and expansion from Cu and Si targets irradiated by the fundamental wavelength of a 5 ns Nd:YAG laser in vacuum are examined using time-resolved ICCD imaging and optical emission spectroscopy under multi-spot DOE excitation and compared with plasmas generated using a single Gaussian beam. Copper and silicon are selected as representative metallic and non-metallic materials. The study focuses on plasma expansion dynamics, plume morphology, and emission enhancement, demonstrating that DOE-based spatial beam shaping provides a promising route to improving the analytical performance of LIBS.

Furthermore, the results provide insight into plasma dynamics relevant to applications employing DOE-based beam shaping, including laser ablation in liquids for nanoparticle generation [37,40], pulsed laser deposition (PLD), time-of-flight mass spectrometry studies of ion generation, and space propulsion and debris deorbiting concepts [41].

## 2. Experimental Details

The targets used in the experiments were a high-purity (99.999 %) Si (100) wafer (Pi-Kem) and a high-purity (99.99 %) Cu disc. The laser (Ekspla NL230) operated at a wavelength of 1064 nm with a pulse duration of 5 ns and a repetition rate of 5 Hz. The pulse energy was monitored using a pyroelectric energy meter (Ophir PE25BF-DIF-C) and adjusted to achieve the desired fluence. To minimize crater formation and ensure reproducible ablation conditions, the target was mounted on a rotating disc holder with adjustable rotational speed and XY translation. The motion was controlled via a DC motor and a linear translation stage coupled through a vacuum feedthrough.

A diffractive optical element (DOE) beam splitter (HOLO/OR MS 574-J-Y-A, diameter of 25.4 mm, separation angle 0.39°) was used to generate a 2×2 array of laser spots from a single incident Gaussian beam. The DOE consists of a microstructured phase pattern designed to produce multiple parallel diffracted beams with nearly equal intensities [39]. In this configuration, the incident beam was split into sub-beams with equal spot sizes (~533 µm, $1/e^2$). The total input laser energy was adjusted to be four times higher than that used for single-spot irradiation, ensuring that the fluence per spot remained constant at a ~10 J/cm$^2$ peak value. The DOE was combined with a focusing lens (250 mm focal length) to produce an array of four focused spots with a square side of ~ 1.4 mm at the target plane.

The experimental set-up is shown in figure 1. Multiple diagnostics were employed simultaneously to characterize the plasma dynamics, including spectrally, spatially, and temporally resolved plume emission in the visible spectral range, as well as time-resolved ICCD imaging oriented normal to the plasma expansion axis. The plasma emission was collected using two mirrors and a cylindrical lens (f = 75 mm) and directed out of the vacuum chamber (base pressure ~10$^{-5}$ mbar) onto the entrance slit of a spectrometer (Kymera 328i: Oxford Instruments, Andor). The spectrometer features a motorized Czerny-Turner configuration with a four-grating turret, allowing operation at different spectral resolutions. An intermediate grating (1200 lines mm$^{-1}$) with an entrance slit width of ~ 80 µm was used for most measurements. The emitted light was detected using an ICCD camera (iStar DH340T-18U-03 Andor Technology) with a resolution of 2048×512 pixels, thermoelectrically cooled to -30 °C.

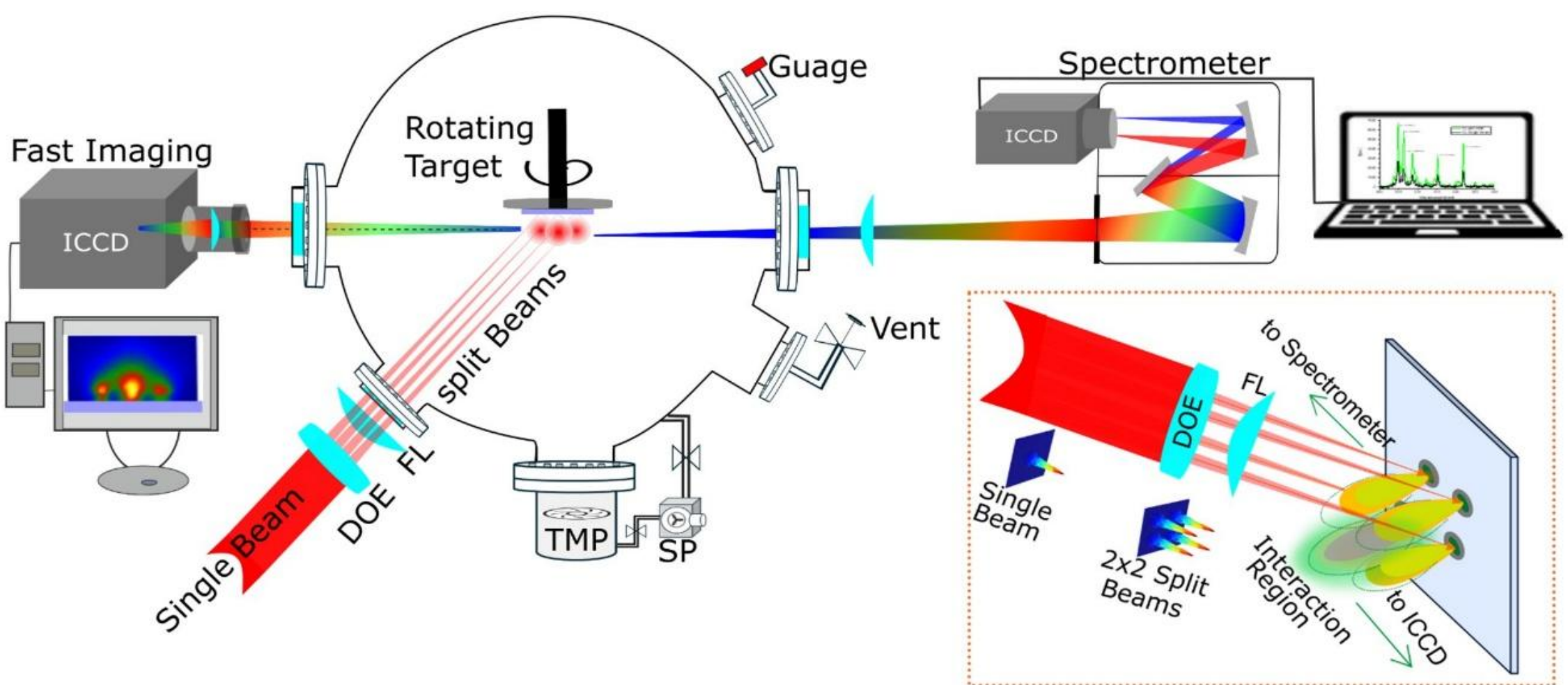

**Figure 1.** Schematic of the LIBS experimental set-up with a diffractive optical element (DOE) for laser beam splitting. The interacting plasma plumes are produced under vacuum (~10$^{-5}$ mbar) maintained by a scroll pump (SP) and a turbomolecular pump (TMP). The laser beam path (red lines) before and after splitting, the focusing lens (FL), spectrometer, and ICCD for time-resolved imaging are indicated. Inset: Illustration of DOE beam splitting and the resulting interaction of the four sub-beams. The ICCD camera is oriented perpendicular to the plasma expansion axis (side view). In this projection, two of the four plasma plumes overlap along the line of sight, resulting in three visible plumes.

Imaging was performed using a second ICCD camera (iStar DH334T-18U-A3, Andor) with a minimum gate width of ~ 3 ns. Following the approach described in [42], the gate width was set to the minimum value plus approximately 20 % of each delay time in order to compensate for the decrease in emission intensity during plasma expansion. The imaging system was equipped with a Navitar 50 mm f/2.8 objective lens with adjustable magnification. For selective imaging of emission from neutrals of Cu plume, a bandpass filter centred at 520 nm (FWHM 10 nm, Thorlabs) was employed.

The spectrometer and both ICCD cameras were externally triggered by a TTL signal from the laser synchronization output. Due to the side-view imaging geometry along the diagonal of the square array, two of the four laser spots overlapped in projection along the line of sight, resulting in three visible plumes in the recorded images.

## 3. Results and Discussion

In this section, the results of spectroscopy and time-resolved imaging are presented via comparing the regimes with the single beam and multi beam (DOE) actions, followed by a brief discussion and summary.

### 3.1. Emission spectroscopy

Figure 2 compares time-integrated emission spectra of Si irradiated using the DOE and a single beam. The spectral region includes both neutral (Si I) and ionic lines (Si II – Si IV). The signal enhancement factor, calculated as the intensity ratio of DOE to single beam, is highest (~9-fold) for the highest peak of the ionic component Si IV at 408.9 nm.

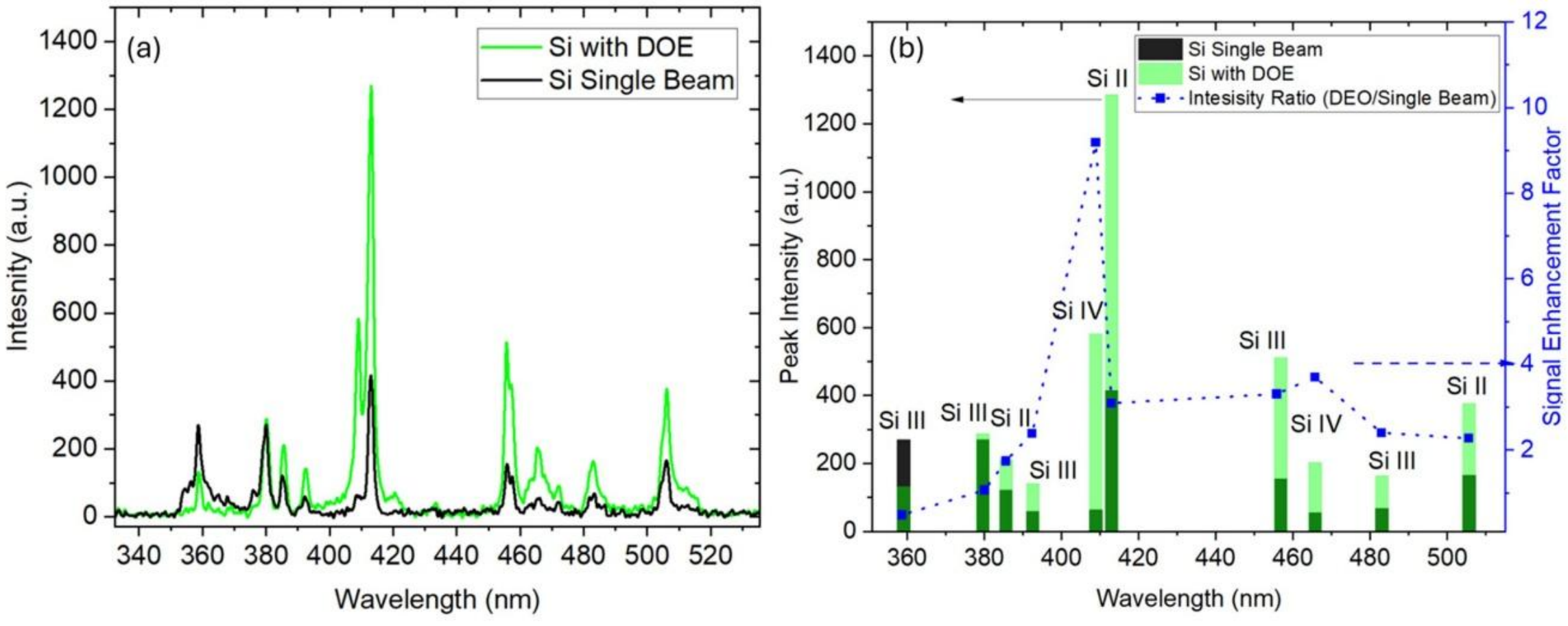

**Figure 2.** (a) Comparison of emission spectra of Si obtained using single-beam irradiation and DOE. The spectra are integrated over a 5-μs time. (b) Peak intensities of selected Si ion emission lines (Si II–Si IV) for both single-beam and DOE configurations, along with the corresponding enhancement factors for DOE irradiation.

Similarly, the spectral emission intensity of Cu irradiated with the DOE exhibits enhancement compared to the single-beam case (figure 3). The DOE increases the population of higher charge states in Cu, which have significantly higher ionization potential (20.29 eV for Cu II compared to 7.7 eV for Cu I). Their presence therefore indicates the formation of a hotter plasma in the DOE case. The maximum signal enhancement of ~3-fold is observed for the higher charge state Cu II at 506 nm.

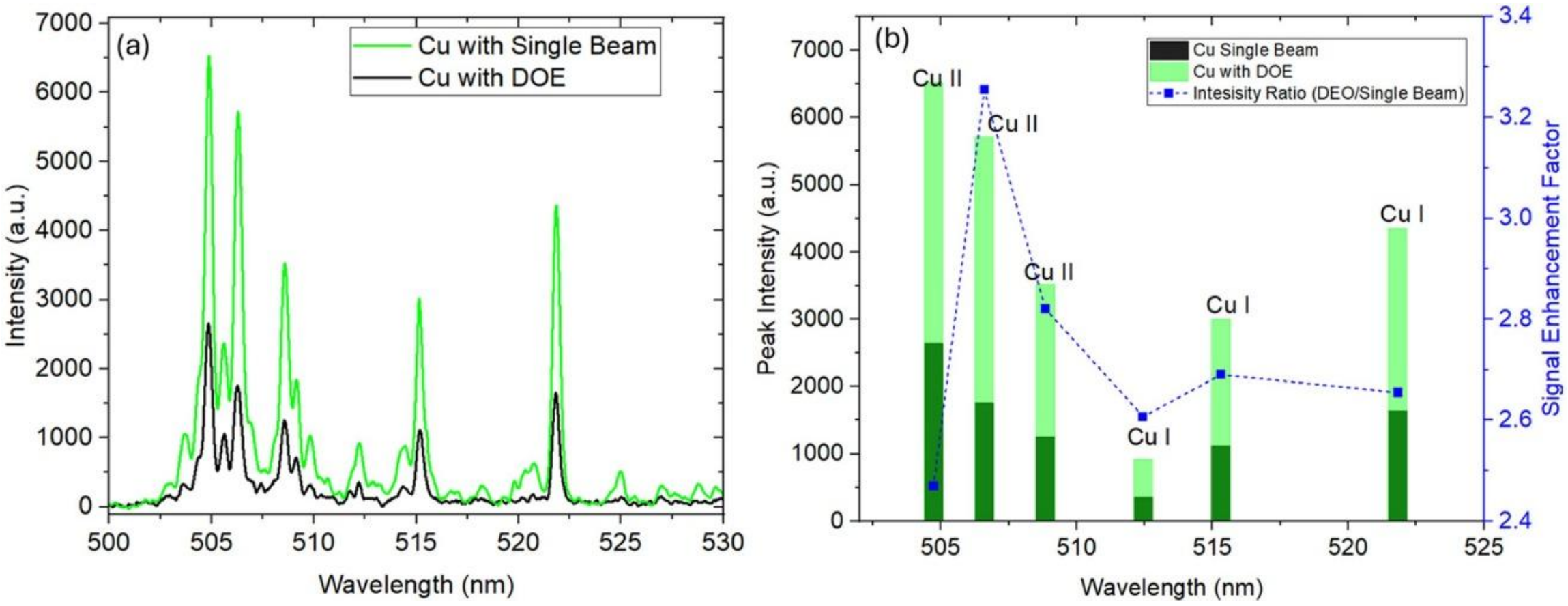

**Figure 3.** (a) Comparison of time-resolved emission spectra of Cu plasma for single-beam irradiation and beam splitting using DOE. The spectra are taken at a 600-ns time delay and include emission lines from both neutral species (Cu I) and singly ionized species (Cu II). (b) Peak intensities of selected Cu spectral lines for both irradiation conditions, together with the corresponding enhancement factors for DOE irradiation.

The observed strong emission lines can be used for estimating plasma opacity by the intensity ratio method. For a homogeneous plasma in local thermodynamic equilibrium (LTE), the intensity ratio of a pair of spectral lines originating from similar neutral or ionic charge states is given by [43]:

$$\frac{I_1}{I_2} = \left(\frac{\lambda_{nm}}{\lambda_{ki}}\right)\frac{A_{ki}g_k}{A_{nm}g_n}\exp[-(E_k - E_n)/k_B T] \tag{1}$$

where $I_1$ and $I_2$ are the line intensities corresponding to transitions $k \to i$ and $n \to m$, respectively; $\lambda$ is the transition wavelength; $A$ is the transition probability; $g$ is the statistical weight; $E$ is the upper-level energy; $k_B$ is the Boltzmann constant; and $T$ is the plasma temperature. By selecting Cu and Si spectral lines with comparable upper energy levels (Table 1), the exponential term can be approximated to unity, simplifying the equation (1) to:

$$\frac{I_1}{I_2} = \frac{\lambda_{nm}A_{ki}g_k}{\lambda_{ki}A_{nm}g_n} \tag{2}$$

This ratio depends only on atomic parameters ($A$, $g$, and $\lambda$), which can be obtained from the NIST atomic database [44].

**Table 1.** Atomic data for selected lines of Cu and Si extracted from the NIST database [44].

| Ion | Wavelength (nm) | Upper-Level Energy (eV) | Statistical Weight | Transition Probability, $\times10^7$ (s$^{-1}$) | Transitions | |
|---|---|---|---|---|---|---|
| | | | | | Upper Level | Lower Level |
| Cu I | 515.32 | 6.191 | 4 | 6.00 | $3d^{10}4d\ ^2D_{3/2}$ | $3d^{10}4p\ ^2P_{1/2}$ |
| Cu I | 521.82 | 6.192 | 6 | 7.50 | $3d^{10}4d\ ^2D_{5/2}$ | $3d^{10}4p\ ^2P_{3/2}$ |
| Si III | 392.45 | 28.55 | 9 | 34.7 | 3s5g 1G | 3s4f 1F° |
| Si III | 482.89 | 28.55 | 11 | 22.5 | 3s 5g 3G | 3s4f 3F |

The Cu spectral lines selected for the intensity ratio analysis are the non-resonant doublets of Cu I at 521.82 nm and 515.32 nm. These lines are particularly suitable due to their nearly identical upper-level energies (6.192 eV and 6.191 eV, respectively). They arise from the electronic transitions $3d^{10}4d\ ^2D_{5/2} \rightarrow 3d^{10}4p\ ^2P_{3/2}$ and $3d^{10}4p\ ^2D_{3/2} \rightarrow 3d^{10}4p\ ^2P_{1/2}$, respectively [44].

Using the atomic parameters listed in Table 1, the theoretical intensity ratio of these lines is calculated to be 1.85. Experimentally, the observed intensity ratios at a delay time of approximately 600 ns after the laser pulse are nearly similar for the two cases, yielding values ~1.44 for the DOE and ~1.47 for the single-beam configuration. At a later delay of 2000 ns, this ratio for the DOE reaches a maximum of ~1.71, corresponding to a relative deviation of ~7.6% from the theoretical prediction, while the plume generated by the single pulse has already lost its luminosity. Using a similar methodology, Unnikrishnan *et al.* [45] reported a maximum experimental ratio of 1.6 for these Cu lines, supporting the assumption of optically thin plasma and the LTE conditions under both single- and multibeam irradiation.

Similarly, for silicon, the ratio of experimental to theoretical intensities for two selected Si III lines (482.9 nm / 392.45 nm) lies within a relative deviation of ~22%. It is worth noting that uncertainties in atomic parameters, particularly transition probabilities, can be as high as ~18% [44], which can partly account for the observed discrepancies.

The electron temperature can be inferred from the intensity ratio of two successive ionization stages of the same element (denoted as $I'$ and $I$), assuming that the population of energy levels follows the Boltzmann distribution [46]. The relationship is given by:

$$\frac{I'}{I} = \frac{f'g'\lambda}{fg\lambda'}\left(\frac{kT}{E_H}\right)^{3/2} \exp[(E - E' - E_\infty)/k_B T] \quad (3)$$

where $I$, $f$, $g$, $\lambda$, $E_H$, $E$, and $E_\infty$ represent the total line intensity, oscillator strength, statistical weight, wavelength, ionization energy of hydrogen, excitation energy, and ionization energy of the lower ionization stage, respectively. Primed quantities refer to the higher ionization stage.

The electron temperature was estimated using the intensity ratio of two successive ionization stages of silicon. The measured ratio of the Si III line at 380.65 nm to the Si II line at 385.60 nm was 1.05 for single-beam irradiation and 1.36 for DOE conditions. Using atomic data from the NIST database and assuming a typical electron density of $N_e \sim 10^{17}$ cm$^{-3}$, these ratios correspond to electron temperatures of approximately 1.55 eV (single beam) and 1.80 eV (DOE). The inferred temperatures show only weak sensitivity to reasonable uncertainties in electron density and atomic parameters and are consistent with previously reported values for silicon laser-produced plasmas under similar experimental conditions [47].

### 3.2. Time-Resolved Imaging

#### 3.2.1. Plume-Plume Interaction

The single-beam plume dynamics exhibits the characteristic hydrodynamic free expansion in vacuum (figure 4(a)). In contrast, under DOE irradiation (see figure 4(b) for Cu and figure 5 for Si), the simultaneously arriving sub-beams generate multiple co-propagating plumes. These plumes expand both axially (normal to the target surface) and radially (along the target surface), leading to plume interaction.

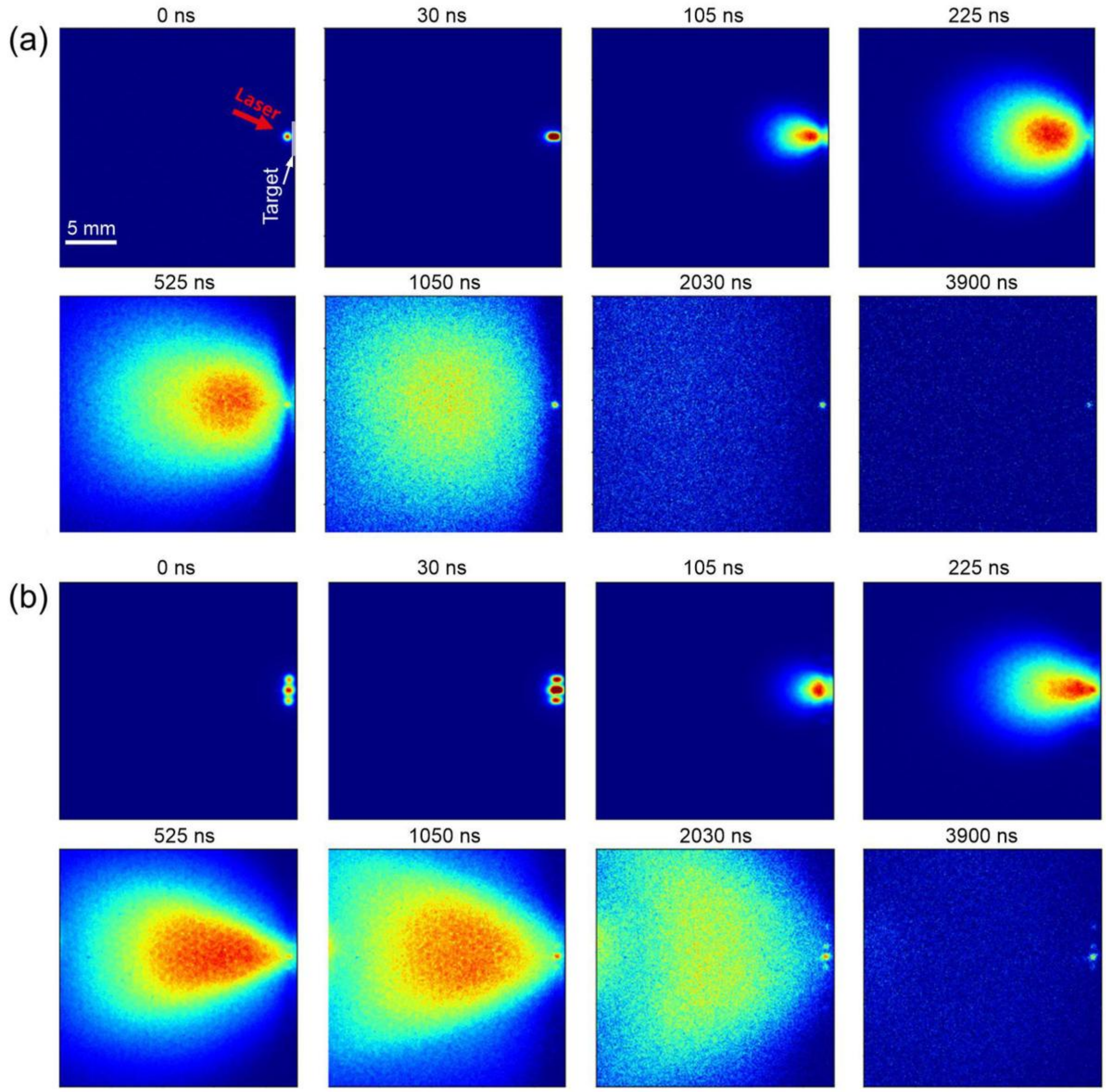

**Figure 4.** Time-resolved ICCD images of plasma generated by irradiation of a Cu target with a Nd:YAG laser (1064 nm, fluence 10 J cm$^{-2}$) under vacuum conditions: (a) single-beam irradiation and (b) multi-beam irradiation using DOE. Plasma expansion (towards the left) was recorded with an ICCD camera using a gate width of 3 ns plus 20% of the delay time. The emission intensities are normalized to their respective maximum values and exposure time for improved visual comparison. In the DOE case, individual Cu plumes progressively interpenetrate and diffuse into one another, forming a more uniform central plasma region with a larger effective emission volume and extended plasma lifetime.

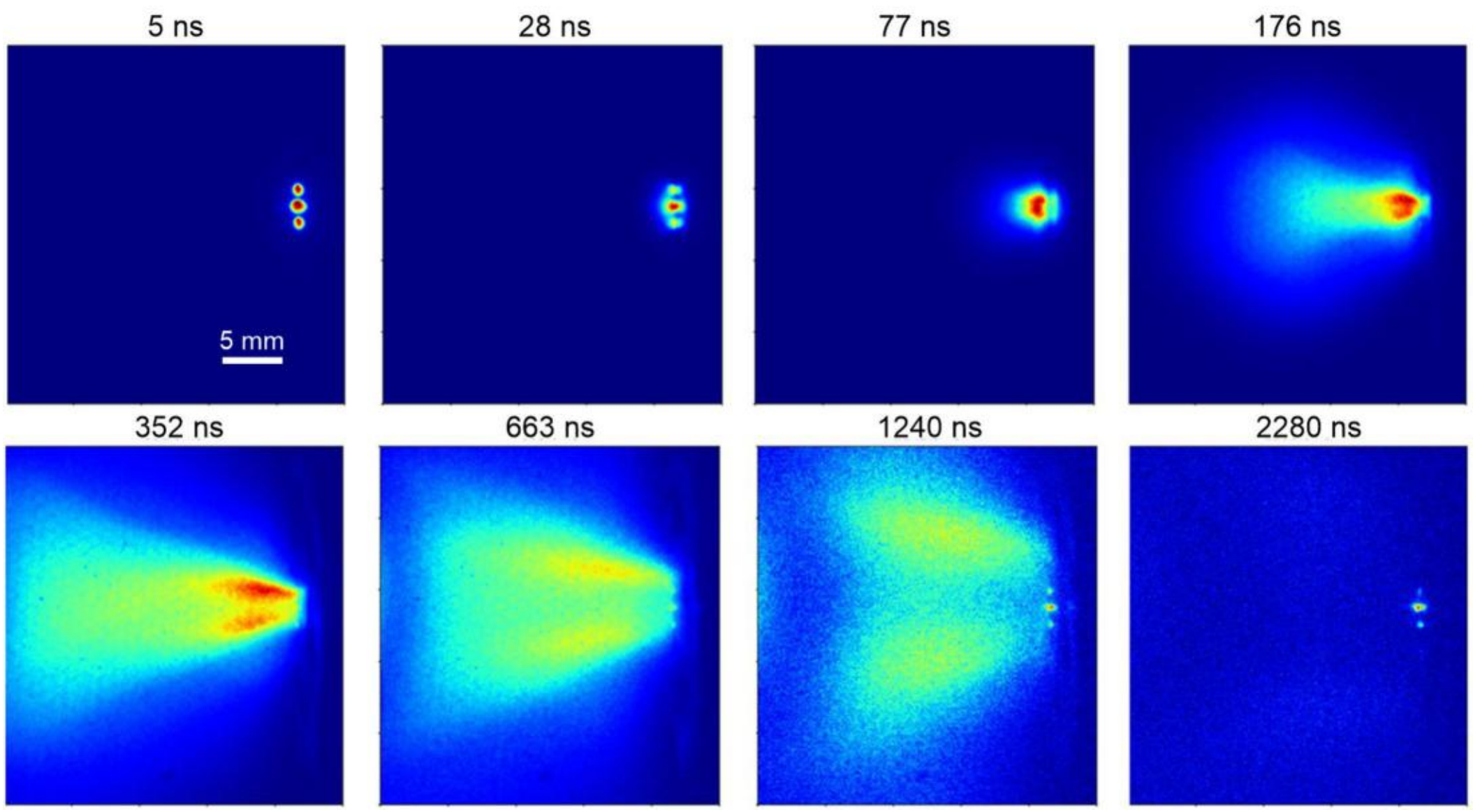

**Figure 5.** Time-resolved ICCD images showing the plasma expansion of Si target irradiated using DOE. The intensities in each image are normalized to their maximum intensity and exposure time for visual clarity. As time increases, the plumes manifest two lobes in the direction normal to the plume expansion axis.

Figure 6(a) presents the plume intensity profiles in the transverse direction (parallel to the target surface). These profiles were extracted from time-resolved 2D ICCD images of the Cu plasma (see figure 4) generated under DOE irradiation, using pixel-by-pixel spatial integration along the direction normal to the target surface at each time delay. As seen in figure 6(a), at an early expansion time (~5 ns), the plasma continues to absorb laser radiation via inverse bremsstrahlung, resulting in additional plasma heating. Consequently, the intensity profile exhibits its maximum emission, dominated by continuum radiation arising from free–free (bremsstrahlung) and free–bound (recombination) transitions [48].

At early times (< 50 ns; figure 6(b)), the plumes generated by the multibeam configuration exhibit expansion behaviour comparable to that of the single-beam case, with an axial velocity of the plume core of ~2.1 km $s^{-1}$ (the plume core is defined by the position of maximum emission). During this stage, the lateral expansion – estimated from the temporal evolution of the intensity in figure 6(a) – yields a transverse plume-edge velocity of ~ 3.85 km $s^{-1}$ (the plume edge is defined at 10% of the maximum intensity). By ~75 ns, the emissions from the individual plumes merge into a single intensity peak, marking the onset of plume interaction and subsequent collective expansion. As the plume continues to expand and undergo radiative and gas-dynamic cooling, the overall emission intensity decreases rapidly.

At later delays, up to ~225 ns, the axial plume velocity exhibits a significant reduction for the DOE configuration compared to the single-beam case. While the single-beam plume expands rapidly (~24.4 km $s^{-1}$), the corresponding velocity in the multibeam configuration is much lower (~9.8 km $s^{-1}$; see inset in figure 6(b)). This deceleration in the DOE case can be attributed to enhanced collisional interactions and stagnation effects arising from the mutual interaction of adjacent plumes, which leads to efficient slowing of the plume core. As a result, part of the kinetic energy of the colliding plasma species (ions and neutrals) is converted into thermal energy.

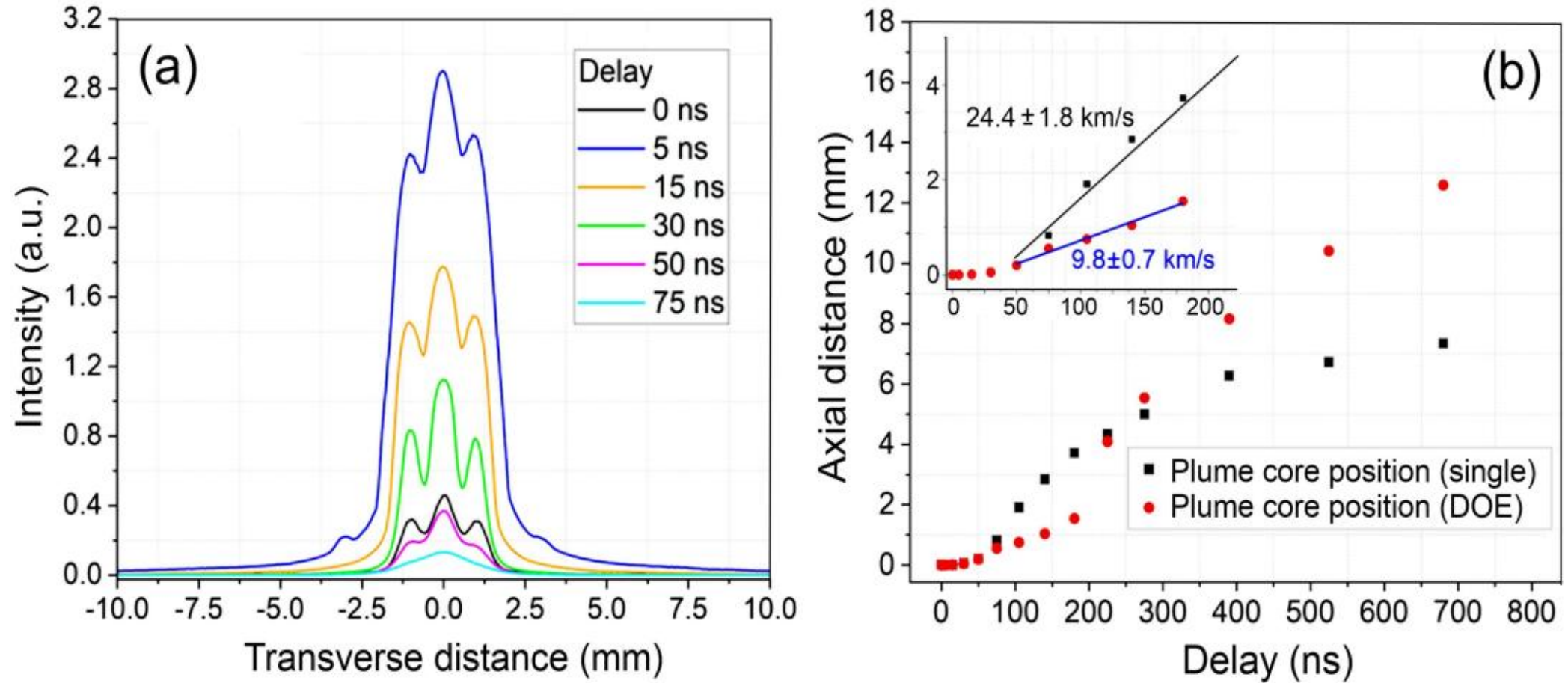

**Figure 6.** (a) Intensity profiles in the direction transverse to the plume axis, extracted from time-resolved 2D ICCD images of Cu plasma generated under DOE irradiation by pixel-by-pixel spatial integration along the direction normal to the target surface for each time delay. At a delay of 5 ns, the profile exhibits the highest intensity, dominated by strong continuum emission during the laser-plasma interaction. The three peaks observed at early times originate from the multibeam plume structure, with the central peak appearing enhanced due to the overlap of two plasma plumes in the side-view geometry. For delay times >75 ns, the profiles merge into a single Gaussian-like distribution. (b) Time evolution of the axial position of the intensity maximum corresponding to the plume core for single-beam and DOE irradiation. The inset shows linear fits of the plume core velocity in the 50–225 ns interval, indicating the early-stage deceleration of the plume for the DOE case (9.8 ± 0.7 km $s^{-1}$) compared to single-beam irradiation (24.4 ± 1.8 km $s^{-1}$), attributable to multibeam plume interaction (collision dynamics).

At later delays, up to ~225 ns, the axial plume velocity exhibits a significant reduction for the DOE configuration compared to the single-beam case. While the single-beam plume expands rapidly (~24.4 km $s^{-1}$), the corresponding velocity in the multibeam configuration is much lower (~9.8 km $s^{-1}$; see inset in figure 6(b)). This deceleration in the DOE case can be attributed to enhanced collisional interactions and stagnation effects arising from the mutual interaction of adjacent plumes, which leads to efficient slowing of the plume core. As a result, part of the kinetic energy of the colliding plasma species (ions and neutrals) is converted into thermal energy.

At even later delays (~225–680 ns), the expansion dynamics reverses: the merged DOE-generated plume expands with a higher axial velocity (core velocity ~21 km $s^{-1}$) compared to the single-beam case (~7.2 km $s^{-1}$). This behaviour can be explained by the increased thermal energy generated through plume–plume interactions, which enhances the pressure-driven expansion. Consequently, the DOE-generated plasma expands more rapidly and remains more emissive than the plasma produced under single-beam irradiation (see the images at 2030 ns in figure 4).

In the case of the Si target irradiated in the DOE configuration (figure 5), the time-resolved images reveal a distinct feature in the plume dynamics that is not observed in the Cu plume expansion (figure 4 (b)). This feature manifests as a transverse splitting in the plume into two lobes, becoming evident at delays from ~170 ns. The corresponding spatial intensity profiles, obtained by pixel-by-pixel integration along the direction normal to the target surface, show that the separation between the intensity maxima of the two lobes increases from ~1.25 mm at ~ 220 ns to ~10 mm at ~1870 ns.

The absence of such lobes in the Cu images can be explained by a lower density of the interacting plumes compared to Si ablation. The ablated mass of Si and Cu under the present conditions (1064 nm, ~10 J $cm^{-2}$, nanosecond pulses) was estimated based on literature data. For Si, extrapolation of ablation-depth measurements reported by Tao *et al*. [49] yields an average ablation depth of ~0.5 μm at 10 J $cm^{-2}$, corresponding to an ablated mass of ~1.2 μg $mm^{-2}$ per pulse. Although the shorter pulse duration used in the present work (~5 ns) implies a higher peak intensity and potentially deeper ablation, the reported values correspond to peak depths at the beam center; thus, ~0.5 μm represents a reasonable estimate for the spot-averaged depth. Based on species analysis by Murakami *et al*. [50], indicating that Si ablation products are dominated by atoms and atomic ions, the total number of emitted Si particles is estimated to be $\sim 2.6\times10^{16}$ atoms per $mm^2$ per pulse.

For Cu, ablation data under comparable conditions are available from Burger *et al*. [51], who reported a removal rate of ~0.5 μg $mm^{-2}$ per pulse at ~10 J $cm^{-2}$. This corresponds to $\sim 5\times10^{15}$ Cu atoms per $mm^2$ per pulse, approximately five times lower than for Si, primarily due to the high reflectivity of Cu in the near-infrared, > 95 % [52]. Even when accounting for uncertainties up to a factor of ~2 arising from extrapolation and spatial averaging, the difference in emitted particle number between Si and Cu remains substantial.

This significantly higher particle density in the Si plume results in stronger transverse pressure gradients within the expanding plasma, which is likely further enhanced by the non-uniform energy deposition in the multibeam configuration. As the plume evolves, these gradients promote lateral expansion, leading to the observable plume splitting. In addition, variations in local density and temperature, together with plume–plume interaction and possible self-absorption effects, may further contribute to the formation and persistence of the two-lobed structure. A more detailed discussion is provided in section 3.3.

### 3.2.2. Spectrally Filtered Fast Imaging

The spectroscopic analysis of the Cu plasma indicates that the spectral window of 480–530 nm contains intense and well-resolved emission lines from both excited neutrals (Cu I) and ionic species (Cu II). This facilitates the use of bandpass (BP) filters for selective imaging of the temporal evolution of excited Cu neutrals in the visible range, where the ICCD camera is highly sensitive. In the present study, a BP filter centred at 520 nm ± 10 nm was employed to isolate emission from neutral Cu lines, primarily at 515.32 nm and 521.82 nm.

Figure 7 presents spatial intensity profiles extracted from time-resolved BP-filtered ICCD images of Cu plasmas for both single-beam and DOE configurations, obtained by integration along the axial and transverse directions (normal and parallel to the target surface, respectively). The axial profiles (figure 7, left column) for both irradiation modes exhibit two distinct components: a slowly expanding, higher-intensity core near the target, and a rapidly propagating, lower-intensity leading front. The front expands along the plume axis with a velocity of ~27.3 ± 0.85 km/s, whereas the near-target core shows only limited motion, with a velocity of ~ 2.0 ± 0.06 km/s.

The axial intensity profiles (figure 7, left column) reveal a consistent temporal evolution of plume emission and dynamics for both single-beam and DOE configurations, characterized by splitting into fast and slow components. This phenomenon of axial plume splitting is well established in laser-produced plasmas [53–58] and has been observed in both vacuum and ambient gas. However, a comprehensive understanding of the underlying mechanism remains an active area of theoretical and experimental research [59,60]. In vacuum, the formation of a self-consistent ambipolar electric field – often described in terms of a dynamic double layer [53,58] – provides a plausible explanation for the differential acceleration of charged species, leading to the coexistence of fast and slow plume components. The neutral species observed in the fast plume component (figure 7) likely originate from the recombination of the accelerated ions during the plasma expansion [56,61,62].

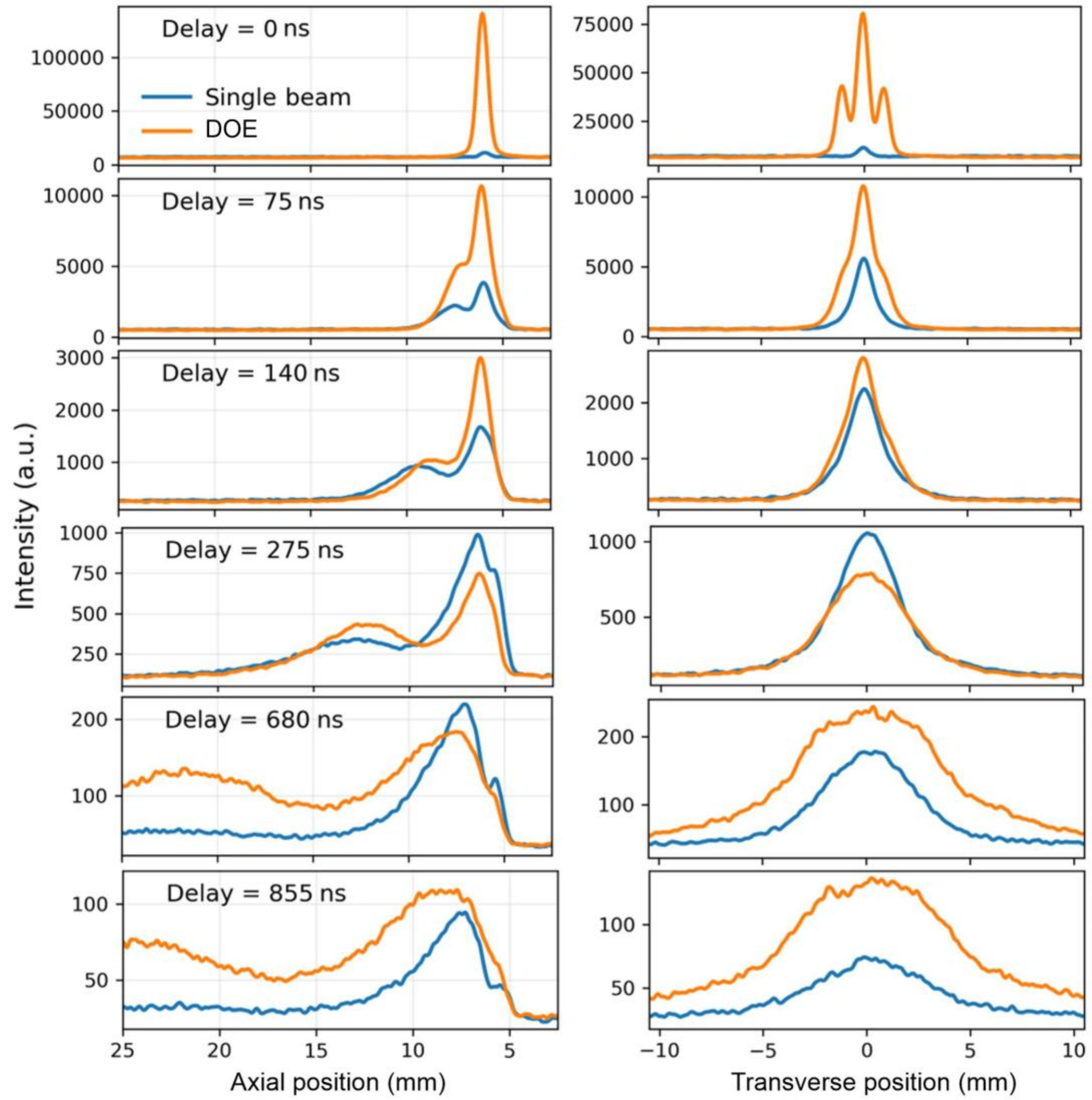

**Figure 7.** Left column: Temporal evolution of axial intensity profiles obtained by integration along the transverse direction. Right column: Transverse intensity profiles obtained by integration along the axial direction (normal to the target surface). In both single-beam and DOE configurations, the profiles were derived from time-resolved ICCD images recorded using a 520 nm BP filter, thereby selectively capturing emission from excited Cu neutrals. In the axial direction, the plume separates into two distinct components: a low-intensity leading front and a high-intensity core. The transverse profiles in the DOE case at early delay times exhibit the multi-lobed structure.

In the present experiments, plume-front splitting is observed at early delay times for both single-beam and DOE irradiation. At ~50–75 ns, both cases exhibit a clear separation into fast and slow components; however, the DOE-generated plume shows stronger emission from excited Cu neutrals, particularly from the slow plume component near the target. As the plume expands, this initially enhanced emission in the DOE case decreases due to radiative losses and recombination, and by ~275 ns the single-beam plume becomes more intense. During intermediate expansion stages (~275–680 ns), the fast component at the plume front generated under the DOE configuration persists for a longer time, remaining visible up to ~680 ns, whereas the corresponding feature in the single-beam case decays more rapidly.

At later delays (~680–855 ns), the emission from the DOE-generated plume again exceeds that of the single-beam case. This sustained brightness can be attributed, as discussed above, to enhanced heating of the ablation products arising from plume–plume interaction prior to the separation of the front from the core, which redistributes energy and supports a higher population of excited neutrals. The front component of the DOE-produced plasma at these stages is strong and comparable in intensity with the core, in contrast to the single-beam configuration, where no fast neutrals are observed (figure 7, left).

Unlike broadband spectral imaging (figures 4 and 5), no small bright spots are observed near the target when using the BP filter for both DOE and single-beam configurations. This indicates that these features are not associated with the emission of neutral atoms. Similar bright spots have been reported at later delays in time-resolved plume imaging employing both nanosecond [63] and femtosecond [64] laser irradiation. In the latter case, they were attributed to a low-velocity population of nanoparticles emitted from the target.

### 3.3. Effect of transverse splitting of the DOE-produced plume

The most significant differences between irradiations in single Gaussian and DOE configurations manifest at the early-to-intermediate stages of plasma expansion, where spatial energy deposition governs both plume morphology and subsequent emission characteristics. Under conventional Gaussian focusing, laser energy is concentrated into a single high-fluence focal volume, resulting in rapid plasma heating and strong pressure gradients along the normal to the target surface. In vacuum, this leads to pronounced axial jetting, rapid plume elongation, and fast rarefaction due to the absence of ambient confinement. Consequently, the luminous plasma volume decreases rapidly, limiting the effective emission lifetime available for LIBS diagnostics. In contrast, multi-spot DOE excitation distributes the incident laser energy across several spatially separated focal spots, each initiating an individual plasma. Each plasma is generated at the same local fluence as in the Gaussian case, their simultaneous formation and spatial proximity fundamentally alter the collective expansion dynamics. For instance, time-resolved ICCD imaging for Cu reveals that plume expansion for the duration of 50–225 ns under multi-spot excitation is characterized by temporary suppression of axial expansion (~10 km/s core velocity) compared to the single beam (~24 km/s). This modified expansion behavior can be attributed to plume–plume interaction effects, as schematically illustrated in figure 8. As adjacent plasmas expand, their lateral pressure fronts intersect, partially redistributing momentum away from the axial direction. This interaction reduces the formation of a single high-velocity axial jet and promotes a more confined plasma volume near the target surface. In vacuum, where confinement is otherwise minimal, such collective effects play a crucial role in slowing plasma rarefaction.

The sustained effective collisional processes within the interacting plasma volume contribute to emission lifetime enhancement. As a result, the plasma remains optically active over an extended time window, which is essential for reliable LIBS measurements in low-pressure environments. The observed emission intensity improvements are directly reflected in the LIBS spectra. For both Cu and Si targets, multi-spot DOE excitation yields stronger atomic and ionic emission lines. The enhancement arises from spatial redistribution of energy and collective plasma effects, which improve excitation conditions without increasing experimental complexity.

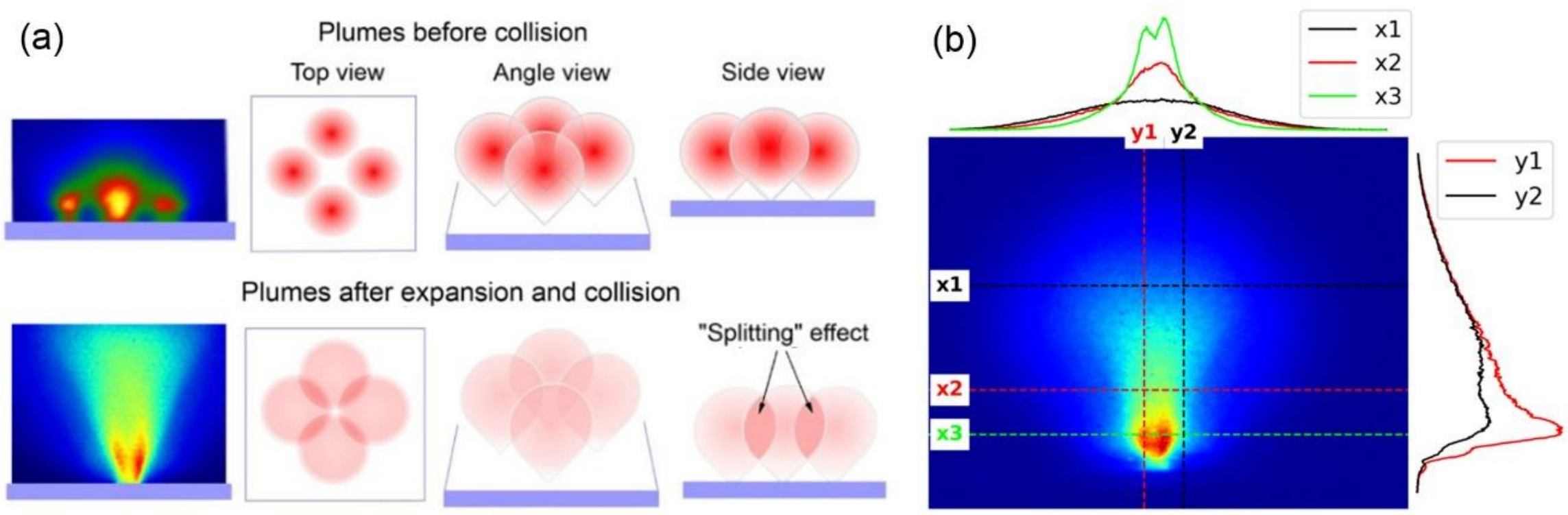

**Figure 8.** (a) Schematic illustration of the interaction of four plasma plumes generated under DOE irradiation before (top panel) and after (bottom panel) interaction. The plumes are shown from top, oblique, and side perspectives. In the experimental configuration, the ICCD camera records the plume in a side view, i.e., perpendicular to the plume expansion axis. In this viewing geometry, the four laser plumes appear as three partially overlapping structures (top panel, right scheme). Upon expansion, the four plumes interact and evolve into a single merged plume in the case of Cu (figure 4, b) whereas for Si they develop a two-lobed structure. as illustrated in the bottom panel. (b) Time-resolved ICCD image of the DOE-generated Si plume at a delay time of 176 ns. The upper panel shows intensity profiles integrated along transverse positions (x1, x2, x3), while the right panel presents intensity profiles integrated along two axial directions (y1 and y2).

Thus, the expansion dynamics of multiple plumes in the DOE irradiation configurations represents a significantly more complex process compared to single-beam irradiation. Our work contributes to resolving key aspects of plasma plume interaction dynamics in multibeam laser ablation, an area of increasing interest in recent literature [65,66]. Beyond enhancing the LIBS signal, the use of DOE has also demonstrated high potential for advanced surface micro- and nanostructuring [67] and has been suggested as an effective approach for applications such as laser propulsion and space debris removal [41]. A detailed understanding a charge-dependent ion flux, kinetic energies of plume species, and the overall dynamics of DOE-generated co-propagating plasma plumes from different target materials may open new avenues for laser-based applications. Furthermore, combining experimental investigations with hydrodynamic modelling will provide deeper insight into the governing mechanisms of this complex expansion process, particularly with respect to improving LIBS diagnostics.

## 4. Conclusions

This work demonstrates that diffractive optical element (DOE)-based splitting of a single Gaussian laser beam into multiple sub-beams provides a simple yet highly effective strategy for enhancing LIBS emission. The results reveal that the observed increase in optical emission intensity of Si and Cu plasmas in vacuum originates from controlled collisional interactions between co-propagating plumes generated in the DOE configuration.

Unlike conventional single-beam irradiation, the multibeam geometry fundamentally alters plume expansion dynamics, leading to modified plume morphology, redistribution of energy within the plasma, and prolonged emission lifetimes. These findings highlight a distinct physical mechanism – plume–plume interaction-driven heating – that directly mitigates the intrinsic limitations of vacuum LIBS, namely weak emission and rapid plasma decay.

In addition to performance enhancement, DOE represents a robust and practical beam-shaping solution, offering low alignment sensitivity, reduced optomechanical complexity, compact implementation, and high damage threshold. The ability to tailor plasma dynamics through controlled multibeam interaction opens new opportunities for optimizing LIBS performance and for extending laser-matter interaction strategies.

Overall, DOE-enabled beam splitting provides a versatile platform for controlling plasma properties and represents a promising route toward advanced LIBS systems operating under challenging conditions, as well as for emerging applications in laser processing, propulsion, and debris mitigation.

## Acknowledgements

G.A. Beyene acknowledges support from the Marie Skłodowska-Curie Actions COFUND programme, co-funded by the European Union (Physics for Future, Grant Agreement No. 101081515). I. Mirza, O. Gatsa, N.M. Bulgakova and A.V. Bulgakov acknowledge financing from the Operational Programme Johannes Amos Comenius, co-funded by the European Structural and Investment Funds and the Ministry of Education, Youth, and Sports of the Czech Republic (project SENDISO, No. CZ.02.01.01/00/22_008/0004596). I. Mirza, N.M. Bulgakova and A.V. Bulgakov further acknowledge partial support from the STRATOLASER project, funded by the European Innovation Council (EIC) Pathfinder programme (Grant Agreement No. 101223245). J. JJ Nivas acknowledges financial support from the European Union and the Ministry of Education, Youth and Sports of the Czech Republic (MSCA Fellowships CZ, project No. FZUI-CZ.02.01.01/00/22_010/0002906).